\documentclass[aps,prb,twocolumn,superscriptaddress,showpacks]{revtex4-1}
\usepackage{amsfonts}
\usepackage{amsmath}
\usepackage{amssymb}
\usepackage{graphicx}
\usepackage{epstopdf}
\usepackage{color}
\usepackage{bbold}

\renewcommand{\vec}[1]{\boldsymbol{#1}}

\begin{document}

\title{Valley Hall Effect and Nonlocal Transport in Strained Graphene}

\author{Xian-Peng Zhang}
\affiliation{Department of Physics, National Tsing Hua University, Hsinchu 30013, Taiwan}
\author{Chunli Huang}
\affiliation{Department of Physics, National Tsing Hua University, Hsinchu 30013, Taiwan}
\affiliation{
 Division of Physics and Applied Physics, School of Physical and Mathematical Sciences,
Nanyang Technological University, Singapore 637371, Singapore
}
\author{Miguel A. Cazalilla}
\affiliation{Department of Physics, National Tsing Hua University and National
Center for Theoretical Sciences (NCTS), Hsinchu 30013, Taiwan}
\affiliation{Donostia International Physics Center (DIPC), Manuel de Lardizabal, 4. 20018, San Sebastian, Spain}

\begin{abstract}
Graphene subject to high levels of shear strain leads to strong pseudo-magnetic fields resulting in the emergence of Landau levels. Here we show that, with modest levels of strain, graphene can also sustain a classical valley hall effect
(VHE) that can be detected in nonlocal transport measurements. We provide a
theory of the strain-induced VHE starting from the quantum Boltzmann equation.
This allows us to show that, averaging over short-range impurity
configurations destroys  quantum coherence between valleys, leaving the elastic scattering time
and inter-valley scattering rate as the only parameters characterizing the
transport theory. Using the theory,  we compute the nonlocal resistance of a Hall bar device in the diffusive regime. Our theory is also relevant for the study of moderate strain effects in the (nonlocal) transport properties of other two-dimensional materials and van der Walls heterostructures.
\end{abstract}
\pacs{}
\maketitle

\affiliation{Department of Physics, National Tsing Hua University, Hsinchu
30013, Taiwan}

\affiliation{Department of Physics, National Tsing Hua University, Hsinchu
30013, Taiwan}

\affiliation{Department of Physics, National Tsing Hua University and National
Center for Theoretical Sciences (NCTS), Hsinchu 30013, Taiwan}

\affiliation{Donostia International Physics Center (DIPC), Manuel de
Lardizabal, 4. 20018, San Sebastian, Spain}

\affiliation{Department of Physics, National Tsing Hua University and National
Center for Theoretical Sciences (NCTS), Hsinchu 30013, Taiwan}
\affiliation{Donostia International Physics Center (DIPC), Manuel de
Lardizabal 4, 20018 San Sebastian, Spain}

\section{Introduction}

The manipulation of the valley degree of freedom, i.e. the field of
\emph{valleytronics}, is currently under intensive research, not only concerning
graphene~\cite{jiang2013generation,gorbachev2014detecting,Falko1,Falko2,PhysRevB.94.121408} but also other two dimensional
(2D) materials. \cite{xiao2010berry,shimazaki2015generation,sie2015valley,lee2016electrical}
Indeed, the generation of valley currents has been recently demonstrated~%
\cite{gorbachev2014detecting} in graphene devices deposited on a Boron
Nitride (hBN) substrate. The effect of the  hBN substrate is to break the symmetry between the two
sublattices of the honeycomb lattice, which opens an energy gap at the (Dirac) point
where the conduction and valence bands meet.~\cite%
{neto2009electronic,katsnelson2012graphene} As a result, a finite Berry
curvature, with opposite sign at opposite valleys, endows electrons
with an anomalous velocity and leads to a valley-polarized current in the bulk  transverse to
the applied electric field.~\cite{Haldane_Berry_Curvature_FS,xiao2010berry} This phenomenon, known as the valley Hall
effect (VHE), can be detected as large enhancement of the nonlocal resistance in a Hall bar device.~\cite{gorbachev2014detecting,abanin2009nonlocal,PhysRevB.94.121408}

Here, we report on a different approach to generate valley-polarized currents in
graphene. Since strain can be controlled more easily
than the magnitude of the hBN-induced gap,  it will allow for a larger
tunability of the effect, thus providing a novel link between  valleytronics and \emph{straintronics}.~\cite{guinea2010energy,vozmediano2010gauge,vozmediano_gauge2,Falko1,Falko2, roche2016quantum} Furthermore, strain also provides a ``dual
counterpart'' to the VHE emerging from Berry curvature in momentum space.~%
\cite{xiao2010berry,gorbachev2014detecting} This is because in graphene and other 2D materials~\cite{Cazalilla_Strain_TMDCs,Pearce_TMDCs}
strain can be described as a (pseudo) gauge field, which induces a
(Aharonov-Bohm-like) phase in real space.

A direct consequence of the strain-induced gauge fields is emergence of pseudo-Landau levels, whose experimental observation has been reported in both real~\cite%
{Levy544,shioya_uniaxial_strain,Li_valley_polarized_pLL} and artificial
graphene systems.~\cite{manoharan,mordecai} Nevertheless, the observation of
quantized valley edge currents (i.e. the quantum VHE), which was predicted
in Ref.~\onlinecite{guinea2010energy}, has not yet been reported. Indeed, the
requirements for the latter are rather stringent, involving devices under
relatively high shear strain, low temperatures, and high mobility graphene
which is free of atomic-size defects and armchair-like~\cite{Falko2} edges. On the other hand, bulk valley Hall currents can be generated in graphene nanoresonators by the application of pulsed strain, as predicted in Ref.~\onlinecite{jiang2013generation}. However, the  valley currents that are
discussed below do not require either pulsed strain or highly strained,
high-mobility devices. The strain-induced VHE that we predict should be observable with fairly modest strain levels in hall bar devices. Furthermore, unlike recent work along similar lines,~\cite{peeters_recent,Sandler_strain,jiang2013generation,Falko1,Falko2} which focuses on nanometer-size devices and ballistic transport,  our
results apply to much larger and disordered devices in the micrometer scale, where conduction takes places in the diffusive regime. The latter are also potentially much more interesting from the application point of view.

 The hallmark of the strain-induced VHE  is the emergence of a large nonlocal resistance in Hall bar devices.~\cite{gorbachev2014detecting,PhysRevB.94.121408} The nonlocal resistance can be computed
 from the diffusion equations for the valley polarization.
Extending previous treatments of the VHE,~\cite{gorbachev2014detecting,abanin2009nonlocal,PhysRevB.94.121408} which have relied on a phenomenological treatment of the  diffusion equations, here
we provide a microscopic derivation of the diffusion
equations starting from the linearized quantum Boltzmann equation derived in Ref.~\onlinecite{chunli2016graphene}. The latter allows us to account for the full quantum coherence of the valley (pseudo-spin) degree of freedom. We are thus able to
show that, upon averaging over all the possible equilibrium impurity configurations, the diffusion equations depend on only two scattering rates:
the inverse of the mean scattering time  and the inter-valley scattering rate. For the latter, we provide expressions that can be used to extract the scattering rates from first principle calculations of a single impurity potential.

 Finally, it is worth mentioning that the strain-induced valley Hall currents predicted here are neutral currents that do not couple to external magnetic fields. Therefore, unlike  spin currents,~\cite{abanin2009nonlocal,balakrishnan2014giant} valley currents will not display Hanle precession (i.e. modulation of the nonlocal resistance as a function
of the strength of the in-plane magnetic field). Thus,
our findings are relevant for the interpretation of some of the nonlocal transport measurements in graphene decorated with hydrogen~\cite{kaverzin2015electron} and gold adatoms~\cite{neutral2015wang}, for which Hanle precession was not observed.
 Indeed, there is no experimental evidence that the devices studied in  Refs.~\onlinecite{kaverzin2015electron,neutral2015wang} are not 
 subjected to  nonuniform strain.~\cite{private}
 However, the application
of the present theory to such experiments, as well as the study of the interplay with other neutral currents, is beyond the scope
of this work and will be explored elsewhere.~\cite{Zhang2016}

 The rest of the article is organized as follows. In the following section, we describe the   details of the model as well as its validity regime. In Sec.~\ref{sec:linres}, we compute the linear  response of a strained graphene
 and, in particular, the doping and temperature dependence of the valley Hall conductivity. The derivation of the diffusion equation for the valley polarization is provided in Sec.~\ref{sec:diff}. In Sec.~\ref{sec:nonloc}, we
 compute the nonlocal resistance of a Hall bard device, which provides a convenient way to detect the VHE.
In Sec.~\ref{sec:sum} we provide a short summary of our results. Finally, some detailed mathematical expressions are relegated to the Appendix.

\section{Model}

Semiclassically, the electron motion in non-uniformly strained graphene is
described using the following set of equations:
\begin{equation}
\dot{\vec{r}} = \vec{u}_k,\quad \dot{\vec{k}} = \left( e\vec{E} + \tau_z
\dot{\vec{r}}\times \vec{\mathcal{B}}_s\right),  \label{eq:sceq}
\end{equation}
where $\vec{r}$ and $\vec{k}$ are the average position and momentum of a
narrow wave packet of Bloch states, $\epsilon_k = \lambda v_F |\vec{k}|$ the
electron dispersion ($\lambda = +1$ for the conduction and $\lambda = -1$
for the valence band, respectively), and $\vec{u}_k = \vec{\nabla}%
_k\epsilon_k = \lambda v_F \vec{k}/|\vec{k}|$ the carrier group velocity  (henceforth we set $\hbar =1$). In addition, $\vec{E}$ is the applied
electric field, $e < 0$ the electron charge, and $\tau_z \vec{\mathcal{B}}_s$
is strain-induced pseudo-magnetic field.~\cite{guinea2010energy,vozmediano2010gauge,vozmediano_gauge2,katsnelson2012graphene} Note that,
because strain does not break time-reversal invariance (unlike a real
magnetic field), the sign of the magnetic field is opposite at opposite
valleys. In terms of the strain tensor~\cite{vozmediano2010gauge,vozmediano_gauge2,katsnelson2012graphene} $u_{\alpha\beta}$, $\vec{\mathcal{B}}_s= \nabla \times \vec{\mathcal{A}}_s$ where $\vec{\mathcal{A}}_s = \tfrac{\beta}{a}\left(u_{xx}-u_{yy},-2 u_{xy} \right)$ is the pseudo gauge field. Here  $a =
1.42\:\mathrm{\mathring{A}}$ is the carbon-carbon distance and~\cite{guinea2010energy} $\beta
\simeq 2$. In the absence of an electric field
(i.e. $\vec{E} =0$), Eq. \eqref{eq:sceq} predicts that a wave packet of
mean momentum $\vec{k}_0 \neq 0$ moves in a circular orbit and in opposite
directions depending on whether $\vec{k}_0$ lies closer to the $K$ or $%
K^{\prime}$ valley. Such a valley-dependent circular motion of electron wavepackets has been observed numerically.~\cite{Peeters2012}

When quantized, the circular orbits lead to pseudo-Landau levels~\cite%
{katsnelson2012graphene,guinea2010energy, roche2016quantum} (pLLs) with energy dispersion $%
\varepsilon _{n}=\pm \Omega _{c}\sqrt{n}$,
where $\Omega _{c}=\sqrt{2v_{F}^{2}|\mathcal{B}_{s}|}$ is the cyclotron
frequency of graphene. In this work, however, we will explore the
semiclassical regime, for which pLL are absent due to the broadening induced
by disorder and/or temperature ($T$). This is the case when the distance
between consecutive Landau levels, i.e. $\Delta _{n}=\varepsilon
_{n+1}-\varepsilon _{n}$, is smaller or comparable to $\min \{k_{B}T,\tau
_{D}^{-1}\}$, where $\tau _{D}^{-1}$ is the impurity scattering rate (see
below). For large pLL filling factor, i.e. for $\mu \gg \Omega _{c}$, where $\mu=v_{F}k_{F}$ is the Fermi energy (at $T = 0$) and $k_{F}$ the Fermi
momentum, $\Delta _{n}\simeq \Omega _{c} n^{-1/2}$. Taking into account that
$\sqrt{n}\simeq \mu/\Omega _{c}$, the condition $\Delta _{n}\tau
_{D}\lesssim 1$ translates into $\omega _{c}\tau _{D}\lesssim 1$, where $%
\omega _{c}=\Omega _{c}^{2}/\mu =v_{F}|e\mathcal{B}_{s}|/k_{F}$.
Below, we shall see that the modified cyclotron frequency $\omega _{c}$
naturally emerges when the Boltzmann kinetic equation is applied to describe
doped graphene. Besides the low pseudo-magnetic field (i.e low strain)
limit, our results are also applicable in high field limit where $\omega _{c}\tau
_{D}\gg 1$ provided the temperature $T\gg \omega _{c}/k_B$ (where $k_B$ is Boltzmann's constant).

Under the conditions stated above, we can use the following linearized
Boltzmann equation (BE) to describe doped strained graphene:
\begin{align}
&\partial_t \delta n_{k}+ \vec{\dot{r}}\cdot \nabla _{r}\delta n_{k} + \vec{%
\dot{k}}\cdot \nabla_{k}\left[ n_{k}^{0} +\delta n_k \right]= \mathcal{I}%
[\delta n_{k}],  \label{eq:BE}
\end{align}
where $\delta n_k$ is deviation of the electron distribution from the
equilibrium distribution, i.e. $\delta n_k = n_k - n^0_k$, where $%
n_{k}^{0}=n^{0}(\epsilon _{k}-\mu )$, being $n^{0}(\epsilon)=\left[
e^{\epsilon /k_{B}T}+1\right]^{-1}$ the Fermi-Dirac distribution at
temperature $T$ and chemical potential $\mu$. Note that, in order to
correctly account for the quantum entanglement between the two valleys within the $\vec{k}\cdot\vec{p}$ theory,~\cite{katsnelson2012graphene} $%
\delta n_{\vec{k}}$ must be treated as a $2\times 2$ density matrix acting on the
space of valley pseudo-spinors.

In Eq.~\eqref{eq:BE}, the collision integral $\mathcal{I}\left[ \delta n_{%
\vec{k}}\right]$ describes the effect of disorder. Its form has been derived
in Ref.~\onlinecite{chunli2016graphene}, extending the work of Kohn and Luttinger~\cite%
{KohnLuttingerBTE} in order to account for the effects of disorder on the
electron internal degrees of freedom, such as the valley pseudo-spin. To
leading order in the impurity density, $n_{\mathrm{imp}}$,
\begin{align}
\mathcal{I}[ \delta n_{\vec{k}}] & =2\pi n_{\mathrm{imp}}\sum_{\vec{p}}\delta
(\epsilon_{k}-\epsilon _{p})\left[ T_{kp}^{+}\delta n_{p}T_{p k}^{-}\right.
\notag \\
& \qquad \left. -\frac{1}{2}\left\{ \delta n_{k}T_{kp}^{+}T_{ p k}^{-}+T_{k
p}^{+}T_{p k}^{-}\delta n_{k}\right\} \right] ,
\end{align}
where $T^{\pm}_{kp}$ is the scattering matrix for a single impurity (the
system area is assumed to be unity).

At low temperatures, the dominant mechanism that limits the diffusion of
bulk valley currents is the inter-valley scattering caused by atomic-size
impurities and defects. Here we consider a random ensemble of atomic-size
impurities, which are assumed to reside on the
honeycomb lattice sites (e.g. vacancies). Our considerations can be generalized to the other types of impurity potentials classified on
symmetry grounds in Ref.~\onlinecite{cheianov2009ordered}. The effect of
random strain fluctuations, which dominate transport in high-quality devices on substrates like hBN,  has been studied elsewhere,~\cite{Couto_random_strain} and will be neglected here.  Within the $\vec{k%
}\cdot \vec{p}$ theory, the potential for one such impurity
takes the following form:~\cite{basko2008resonant,
cheianov2009ordered,Cheianov_transport_2011}
\begin{align}
V(\vec{r}) &= \left[ v_0 \mathbb{1} + s v_z \sigma_z \right]\delta(\vec{r})
\notag \\
& \quad \left. + v_{xy}\left(\mathbb{1}+s\sigma_z\right) \left(u_x \tau_x +
i u_y \sigma_z \tau_y \right) \right]\delta(\vec{r}),\label{eq:pot}
\end{align}
where the Pauli matrices $\sigma_{\alpha}$ and $\tau_{\alpha}$ ($%
\alpha=x,y,z $ describe the sublattice and valley pseudo-spin, respectively.
In the above expression, the terms in the first line ($\propto v_0, v_z$)
conserve the valley pseudo-spin $\tau_z$ while the terms in the second
line induce inter-valley scattering. The Ising variable $s = +1$ ($s=-1$)
when the impurity sits on the A (B) sublattice. The vector $\vec{u} =
(u_x,u_y) \in S= \{(1,0), (-\tfrac{1}{2}, \tfrac{\sqrt{3}}{2}), (-\tfrac{1}{2%
}, -\tfrac{\sqrt{3}}{2}) \}$ parametrizes the inter-valley scattering potential.~\cite{basko2008resonant, cheianov2009ordered,Cheianov_transport_2011} The
impurities are assumed to form a completely disordered ensemble, which is
the most stable configuration at high doping and temperatures of interest here.~\cite{cheianov2009ordered,Cheianov_transport_2011} Thus, the
configurational variables $(s_{l} =\pm 1,\vec{u}_l \in S)$ can take all the
six possible values allowed by symmetry with equal probability. Hence, upon
solving the scattering problem, the band-projected (on shell) T-matrix can
be obtained, and it takes the general form,
$T^{+}_{kp} = A_{kp} \mathbb{1} + \vec{B}_{kp}\cdot \vec{\tau}$,
where $\vec{B}_{kp} = \vec{B}^{\parallel}_{kp} + \vec{\hat{z}}\: B^{\perp}_{kp}$,
describes the valley-dependent scattering with $\vec{\hat{z}}\cdot \vec{B}^{\parallel}_{kp} = 0$, and
\begin{align}
A_{kp} &= \gamma_0(k) \cos \frac{\theta}{2}, \\
B^{\perp}_{kp} &= i s \tau_z \gamma_z(k) \sin\frac{\theta}{2}, \\
\vec{B}^{\parallel}_{kp} &= \lambda \gamma_{xy}(k) \left[s \left( u_x \cos
\frac{\phi}{2} + u_y \sin \frac{\phi}{2} \right) \vec{\hat{x}} \right.
\notag \\
& \left. \qquad \qquad \qquad + \left(-u_x \sin \frac{\phi}{2} + u_y \cos
\frac{\phi_{kp}}{2} \right) \vec{\hat{y}}\right],
\end{align}
where $\theta = \varphi_{k} -\varphi_{p}$ and $\phi =
\varphi_{k}+\varphi_{p} $, and $\varphi_k = \tan^{-1}(k_y/k_x)$. The
functions $\gamma_{0}(k),\gamma_z(k)$ and $\gamma_{xy}(k)$ depend on $k = |%
\vec{k}|$ (where $\vec{k}$ is the momentum of the incoming electron) and the
potential parameters $v_0, v_z, v_{xy}$ (see e.g. Refs.~\onlinecite{basko2008resonant,hy2015extrinsic} for details of such scattering calculations). However, the important point to notice is that
while $B^{\perp}_{kp}$ depends linearly on the Ising variable $s$, $\vec{B}
^{\parallel}_{kp}$ depends both on $s$ and $u_x,u_y$, in a way such that $%
\overline{B^{\perp}_{kp} \vec{B}^{\parallel}_{kp}} = \vec{0}$, where $%
\overline{O}$ stands for average over the impurity configurations.

\section{Linear Response}\label{sec:linres}

 In order to obtain the response of the system, we parametrize $\delta n_{k}=\rho _{k}\mathbb{1}+
\vec{ \mathcal{P}}_{k}\cdot \vec{\tau}$, where $\rho _{k}$ describes the
charge fluctuations and $\vec{\mathcal{P}}_{k}=\vec{\mathcal{P}}%
_{k}^{\parallel }+\vec{\hat{z}}\: \mathcal{P}_{k}$ (with $\vec{\hat{z}}\cdot
\vec{\mathcal{P}}_{k}^{\parallel }=0$) the valley pseudo-spin fluctuations
of the electron distribution, $\delta n_{k}$.
Summing over the impurity configurations, the
kinetic equations for $\rho_k$, $\mathcal{P}_v$ on one side, and $\vec{
\mathcal{P}}^{\perp}_v$, on the other side, decouple. Thus, in what follows,
we focus on the equations for $\rho_k$, $\mathcal{P}_v$, which describe the
valley Hall effect of interest here.
In addition, the collision integral for the latter is found to be parametrized by two scattering rates: the Drude scattering rate $\tau^{-1}_D = \frac{n_{\mathrm{imp}}k}{4 v_F}( |\gamma_0|^2 + 3|\gamma_z|^2
+ 4 |\gamma_{xy}|^2)$ , and the inter-valley scattering rate $\tau_v =
\frac{n_{\mathrm{imp}}k}{4 v_F}( 8 |\gamma_{xy}|^2)$.

\begin{figure}[t]
\centering
\includegraphics[width=0.5\textwidth]{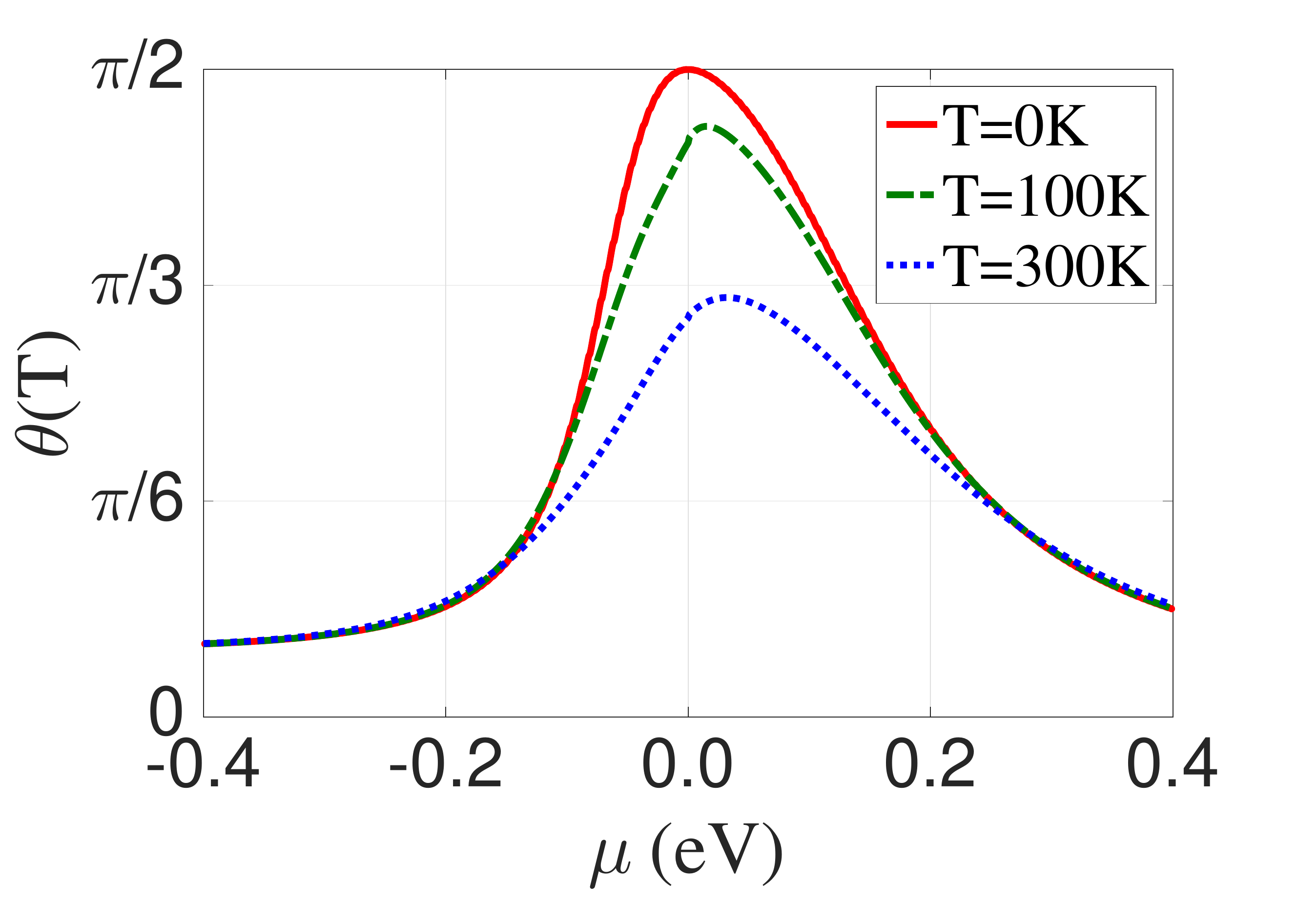}
\caption{(color online) The valley Hall angle, $\theta(T)$ is plotted
against chemical potential $\mu$ for different temperatures $%
T=0,100,300$K. The parameters used  are: $\mathcal{B}_s=0.3\,$ T, $n_{\mathrm{imp}}=5.0\times 10^{11}$ cm$^{-2}$, the impurity (vacancy) potential (cf. Eq.~\ref{eq:pot}) is parametrized by $v_0 =100$ eV, $%
v_{\mathrm{z}}=v_{\mathrm{xy}}=3$ eV, impurity radius $R=0.142$ nm, and cut-off momentum $k_{c}=1 \times 10^{-9}$
m$^{-1}$.}
\label{VHA}
\end{figure}

In the steady state, we employ
the ansatz \cite{chunli2016graphene} $\rho _{k}=\left( \delta \mu +\vec{u}_{c}\cdot \vec{k}\right)
\left[- \partial _{\mu }n_{k}^{0}\right]$ and $\vec{\mathcal{P}}_{k}=\left[ h_{v}
\vec{\hat{
n}}_{0}+\vec{\hat{n}}_{1}\left( \vec{u}_{v}\cdot \vec{k}\right) \right] \left[- \partial _{\mu }n_{k}^{0}\right]$,
which allows us to obtain the constitutive relations by multiplying Eq.~%
\eqref{eq:BE} by $\vec{u}_{k} (\mathbb{1},\tau _{z})$, tracing over $\vec{k},\lambda$ and valley pseudo-spin. Thus,
\begin{equation}
\vec{J} =-\mathcal{D}\vec{\nabla}_{r}\delta n(\vec{r})+\omega _{c}\tau
_{D}\left( \vec{\hat{z}}\times \vec{\mathcal{J}}\right) +\sigma _{D}\vec{E}%
\left( \vec{r}\right),  \label{eq:cr1}
\end{equation}
\begin{equation}
\vec{\mathcal{J}} =-\mathcal{D}\vec{\nabla}_{r}\mathcal{P}(\vec{r})+\omega
_{c}\tau _{D}\left( \vec{\hat{z}}\times \vec{J}\right),  \label{eq:cr2}
\end{equation}
where $\delta n=g_{s}e\mathrm{Tr}\sum_{k}\delta n_{k}$ and $\vec{J}=eg_{s}%
\mathrm{Tr}\sum_{k}\left[ \vec{u}_{k}\delta n_{k}\right] $ are the particle
density and charge current, respectively. $\mathcal{P}=g_{s}e\mathrm{Tr}%
\sum_{k}\left[ \tau _{z}\delta n_{k}\right] $ and $\vec{\mathcal{J}}=eg_{s}%
\mathrm{Tr}\left[ \vec{u}_{k}\tau _{z}\delta n_{k}\right] $ are the valley
polarization and current, respectively ($g_{s}$ is the spin degeneracy). In
the above expression $\sigma _{D}=ne^{2}\tau _{D}/(m_{F})$ ($%
m_{F}=k_{F}/v_{F}$ and $n$ is the carrier density) is the Drude conductivity and $\mathcal{D}=v_{F}^{2}\tau
_{D}/2$ the diffusion coefficient. The last equation describes the classical
VHE, while the second term on the right-hand side of Eq.~\eqref{eq:cr1}
describes the inverse VHE. Next, we solve  Eq.~\eqref{eq:cr1} and \eqref{eq:cr2} for the charge $\vec{J}$ and and valley current $\vec{\mathcal{J}}$, which yields
\begin{align}
\vec{J} &=-\mathcal{D}_{\parallel}\vec{\nabla}_{r}\delta n\left( \vec{r}%
\right) +\mathcal{D}_{\perp} \vec{\hat{z}}\times \vec{\nabla}_r
\mathcal{P}\left( \vec{r}\right) +\sigma _{\parallel}\vec{E}\left( \vec{r}%
\right) ,  \label{JC}\\
\vec{\mathcal{J}} &=-\mathcal{D}_{\parallel}\vec{\nabla}_{r}\mathcal{P}\left( \vec{r}\right) +\mathcal{D}_{\perp} \vec{\hat{z}}\times \vec{\nabla}_r \delta n\left( \vec{r}\right)\notag \\
&\qquad + \sigma _{\perp}\hat{z}\times
\vec{E}\left( \vec{r}\right) ,  \label{JV}
\end{align}
where the longitudinal (transverse) diffusion constant $\mathcal{D}_{\parallel}$ ($\mathcal{D}_{\perp}$) and longitudinal (transverse) conductivity $\sigma_{\parallel}$ ($\sigma_{\perp}$) are given by the following expressions:
\begin{align}
\mathcal{D}_{\parallel} &=\frac{\mathcal{D}}{1+\omega _{c}^{2}\tau _{D}^{2}},
\, \qquad  \mathcal{D}_{\perp} =\frac{\omega _{c}\tau _{D}\mathcal{D}}{1+\omega
_{c}^{2}\tau _{D}^{2}}, \\
\sigma _{\parallel}&=\frac{\sigma _{D}}{1+\omega _{c}^{2}\tau _{D}^{2}},\qquad\sigma _{\perp}=\frac{\omega _{c}\tau _{D}\sigma
_{D}}{1+\omega _{c}^{2}\tau _{D}^{2}}.\label{eq:cond}
\end{align}
Note that, for a uniform electric field, both $\vec{\nabla}_{r}\delta n\left(
\vec{r}\right) $ and $\vec{\nabla}_{r}\mathcal{P}\left( \vec{r}\right) $
vanish, and we obtain the linear response of the system,  $\vec{J}=\sigma _{
\parallel}\vec{E}$ and $\vec{\mathcal{J}}=\sigma _{\perp}\left( \vec{\hat{z}}\times
\vec{E}\right)$. The longitudinal charge conductivity $\sigma _{\parallel} $ is reduced by the strain pseudo-magnetic field in a way  similar to a real magnetic field. Similar to the conventional Hall effect,
both the inverse and direct VHE can be characterized by a figure of merit, namely the \emph{valley Hall
angle} $\theta$, which is defined as follows (see Appendix for the definitions of $\sigma_{\perp}(T),\sigma_{\parallel}(T)$):
\begin{equation}
\tan \theta (T) =\frac{\sigma _{\perp}(T)}{\sigma_{\parallel} (T)},
\end{equation}
At zero temperature,
$\tan \theta (T = 0)=\omega _{c}\tau _{D}$. The dependence of of $\theta (T)$ on the chemical potential $\mu$ at
different temperatures is shown in Fig.~\ref{VHA}. We find that the valley Hall angle can approach $\tfrac{\pi}{2}$ at low doping and low temperatures. This is because the pseudo cyclotron frequency $\omega _{c}$ is inversely proportional to $\mu$ and $\tau_D$ is resonantly enhanced in the neighborhood of the Dirac point.
However, the semiclassical theory becomes less reliable close to the Dirac point where  $\mu =0$. Indeed, 
$\theta  \tfrac{\pi}{2}$ at low doping and $T =0$, which implies that  $\omega_c \tau_D \gg 1$, meaning that the semiclassical theory ceases to be valid, as discussed above. Fortunately, as temperature $T$ is increased, thermal fluctuations
supress the magnitude of $\theta(T)$ at small $\mu$ and we find that a semiclassical regime where $\omega_c \tau_D \gtrsim \theta(T) \gtrsim 1$ also exists at high temperature and low doping.

\section{Diffusion of the valley polarization}\label{sec:diff}

\begin{figure*}[htbp]
\includegraphics[width=0.32\textwidth]{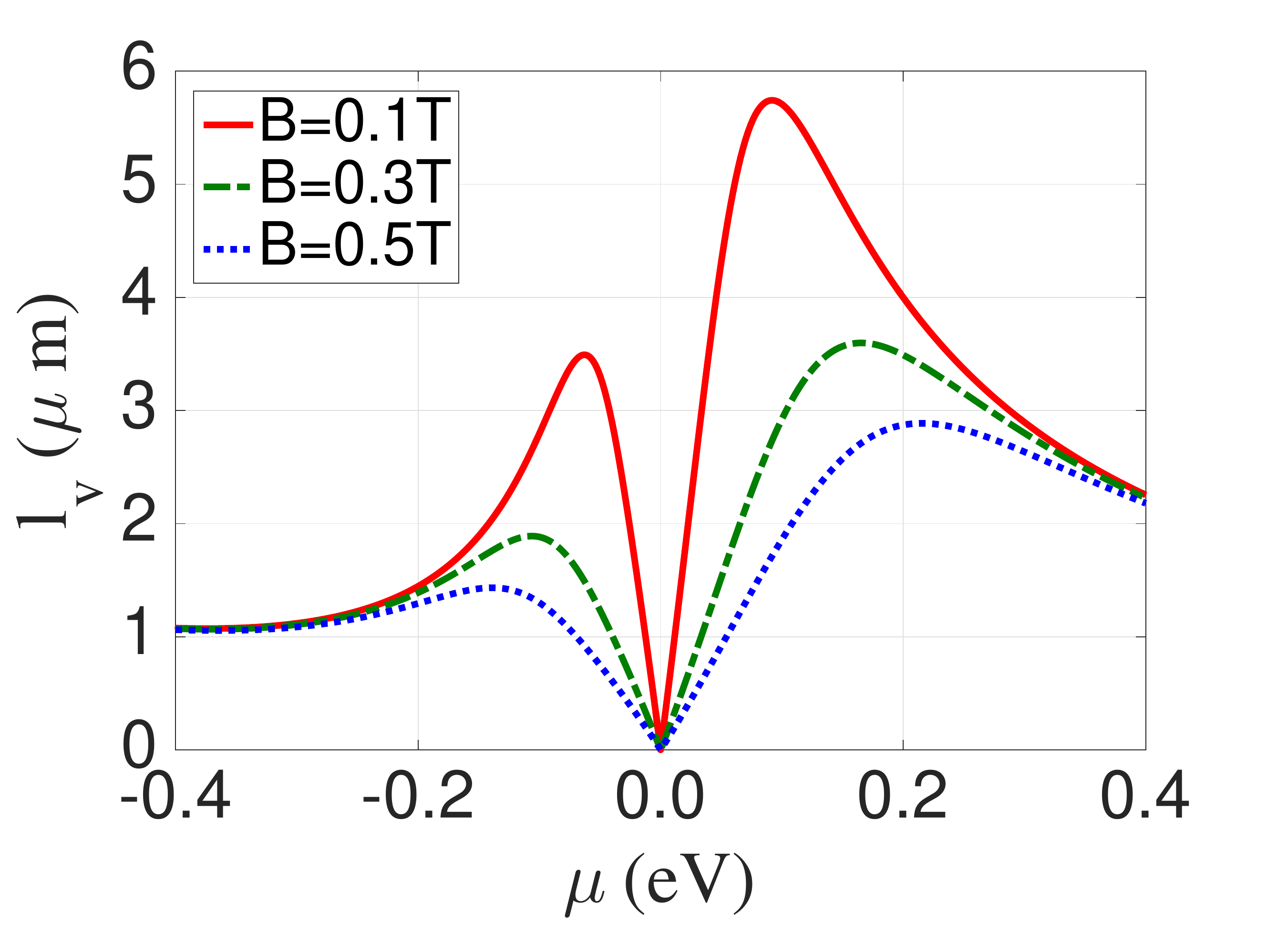} %
\includegraphics[width=0.32\textwidth]{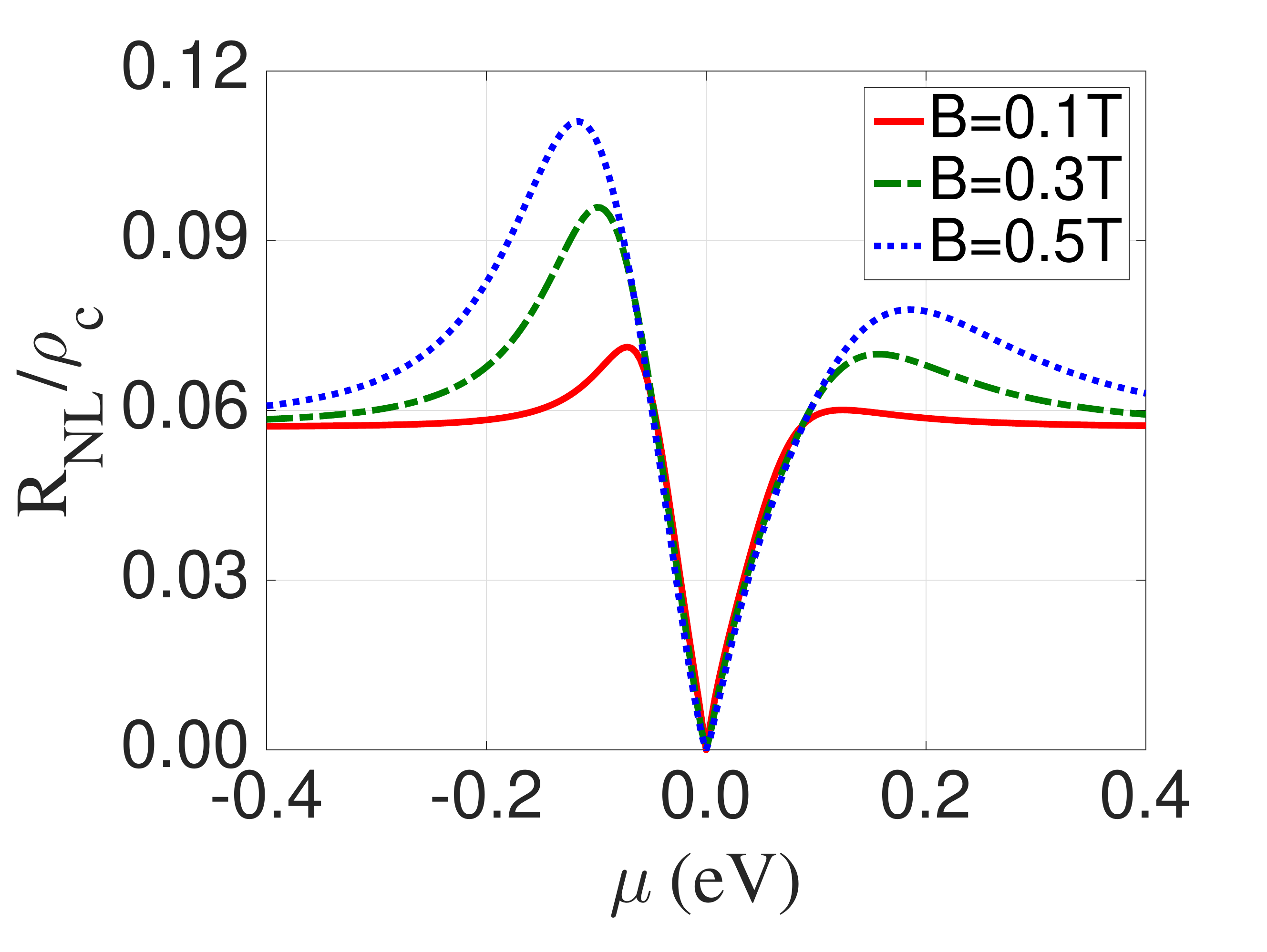}
\includegraphics[width=0.32\textwidth]{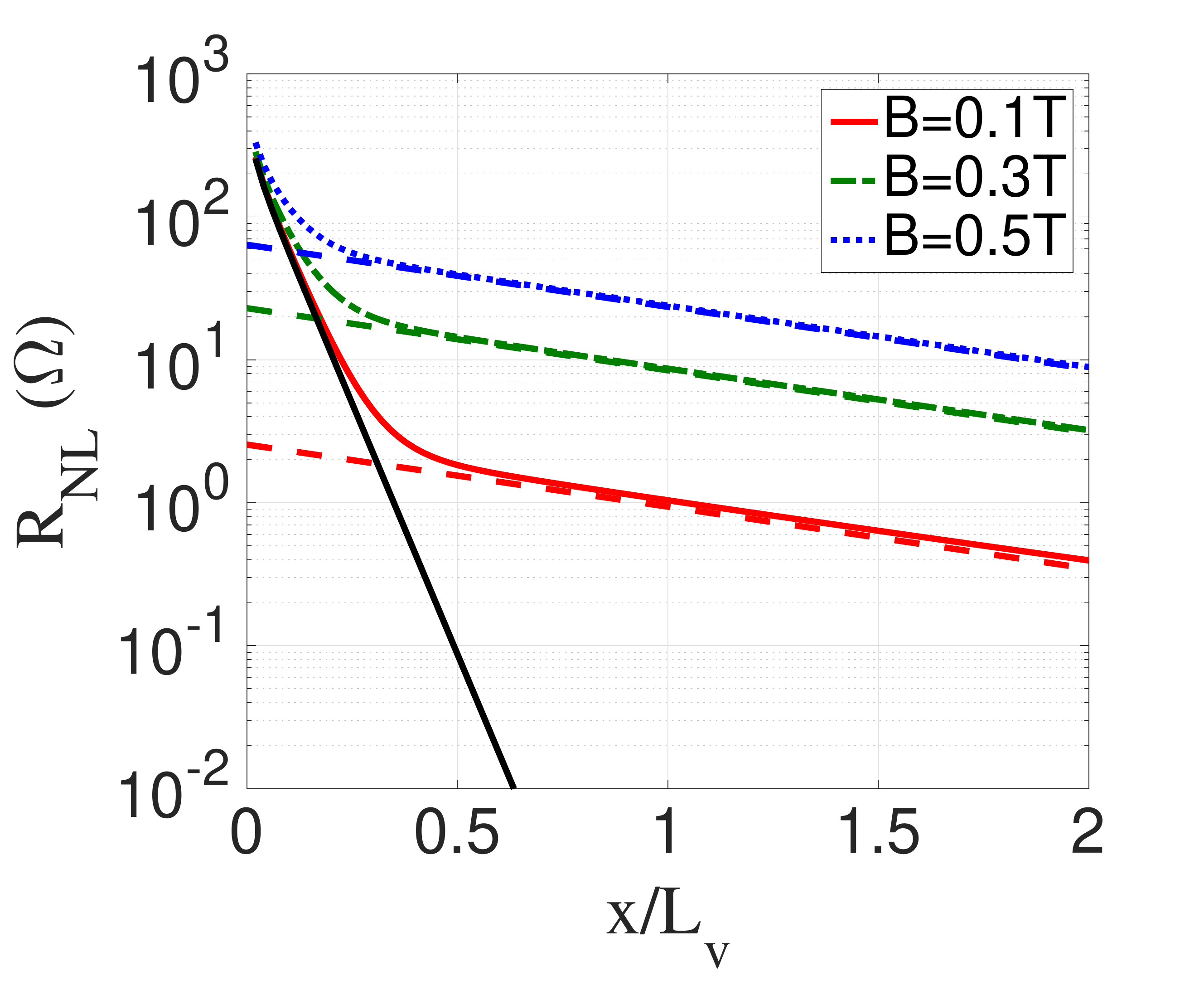}
\caption{(color online) (a) $\ell_{\mathrm{v}}$, in  $\mu\mathrm{m}$, versus the chemical potential $\mu$. (b) Nonlocal resistance $R_{\mathrm{NL}}$ (in units of $\rho_c$) evaluated at $x=1\, \mu\mathrm{m}$ as a function of chemical potential $\protect\mu$. (c) Nonlocal resistance $R_{\mathrm{NL}}$ (in logarithmic scale) as a function of chemical potential $x/L_{\mathrm{v}}$  for fixed chemical potential $\mu=0.15$ eV ($L_{\mathrm{v}}\simeq 5\mu$ m). Note that the decay is controlled by the same length scale $L_{\mathrm{v}}$ for all values of the pseudo-magnetic field. The latter is induced by applying  along the $y$ direction an average (uniaxial) strain of $0.4\%$, $1.2\%$ and $2.0\%$, respectively, to a ribbon of  width $W=1\, \mu\mathrm{m}$. The parameters  are the same as for Fig.~\ref{VHA}.}
\label{NLR}
\end{figure*}

The above quantum Boltzmann equation also allows us to obtain the continuity
equations for the charge and the valley current. After multiplying Eq.~%
\eqref{eq:BE} by $\left( \mathbb{1},\tau _{z}\right) $ and taking the trace
after summing over $\lambda $ and $\vec{k}$, we obtain (in the steady state)
$\vec{\nabla}\cdot \vec{J}=0$, for the charge current and $\vec{\nabla}%
_{r}\cdot \vec{\mathcal{J}}+\mathcal{P}/\tau _{\mathrm{v}}=0$, for the valley
current. By combining the last equation with
Eq.~\eqref{JV}, the diffusion equation for the valley polarization is obtained:
\begin{equation}
\mathcal{D}_{\parallel}\nabla _{r}^{2}\mathcal{P}\left( \vec{r}\right) -%
\frac{\mathcal{P}\left( \vec{r}\right) }{\tau _{\mathrm{v}}}=S\left( \vec{r}%
\right) ,  \label{P6}
\end{equation}
where $S\left( \vec{r}\right) =\vec{\hat{z}}\cdot  \vec{\nabla}_{r}\times \left[
\sigma _{\perp}\left( \vec{r}\right) \vec{E}\left( \vec{r}\right) %
\right] $ is the source of the diffusion. For uniform pseudo-magnetic field, the source term  vanishes everywhere except at the
device boundary where the strain-induced
pseudo-magnetic field vanishes. Eq.~\eqref{P6} indicates the
existence of the following length scale that controls the diffusion of valley polarization:
\begin{equation}
\ell_{\mathrm{v}}=\sqrt{\mathcal{D}_{\parallel}\tau _{\mathrm{v}}}= L_{\mathrm{v}} (1+\omega _{c}^{2}\tau _{D}^{2})^{-1/2}, \label{VDL}
\end{equation}
where $L_{\mathrm{v}}=\sqrt{\mathcal{D}\tau _{\mathrm{v}}}$.
In Fig.~\ref{NLR}(a), we have plotted the length scale
$\ell_{\mathrm{v}}$ against the chemical potential for different values of strength of the pseudo-magnetic field.   We find that the magnitude of $\ell_{\mathrm{v}}$ decreases with the magnitude of the pseudo-magnetic field, as expected from Eq.~\eqref{VDL}. For the present choice of parameters, note that the resulting valley diffusion length $\ell_\mathrm{v}$ (i.e. about $6\, \mu\mathrm{m}$ at $\mu=0.1$ eV) is, in most regimes, larger than the width of the device, $W = 0.5\: \mu\mathrm{m}$. However, as shown in the next section, the decay of the nonlocal resistance along the channel direction is controlled by $L_{\mathrm{v}}$ rather than $\ell_{\mathrm{v}}$.

\section{Nonlocal resistance}\label{sec:nonloc}

Following Beconcini~\emph{et al.},~\cite{PhysRevB.94.121408} we solve
solve the diffusion equation for a Hall bar device geometry,
consisting of a channel of width $W$, which we assume to be infinitely long. Thus,  the solution of the diffusion equations can be found by  imposing suitable boundary conditions (BCs): i) On
the charge current: $J_{y}\left( x,y=\pm W/2\right) =I\delta \left( x\right)
$. This BC describes the current injection (extraction) along the $y$
direction ii)  On the valley current: $\mathcal{J}_{y}\left( x,y=\pm W/2\right) =0$, implying that no valley current flows across the device boundary.

  The solution can be simplified by taking $\delta n \left( \vec{r}\right) \simeq 0$, which amounts to assuming complete screening of the electric
field in the device.~\cite{PhysRevB.94.121408} Thus, the electrostatic potential $\phi (\vec{r})$ obeys the Laplace equation, $\nabla ^{2}\phi (\vec{r})=0$. Using Eq.~\eqref{JC} and (\ref{JV}), the BCs can be recast as:
\begin{align}
I\delta \left( x\right) &=\left[ -\mathcal{D}_{\perp}\partial_{x}\mathcal{P}%
\left( \vec{r}\right) -\sigma _{\parallel}\partial_{y}\phi \left( \vec{r}%
\right) \right] _{y=\pm \frac{W}{2}},  \label{P7}\\
0 &=\left[ -\mathcal{D}_{\parallel}\partial_{y}\mathcal{P}\left( \vec{r}\right)
-\sigma _{\perp}\partial_{x}\phi \left( \vec{r}\right) \right] _{y=\pm
\frac{W}{2}},  \label{P8}
\end{align}
where we have dropped the terms contaning $\delta n\left( \vec{r}%
\right)$. We see that the BCs couple the Laplace equation for
$\phi (\vec{r})$ with the diffusion equation
for $\mathcal{P}\left(\vec{r}\right) $.  Eq.~(\ref{P6}) together with the Laplace equation can be solved using Eq.~(\ref{P7}) and (\ref{P8}) as BCs. Thus, we obtain:
\begin{equation}
\phi \left( \vec{r}\right) =-I\rho_{\mathrm{c}}\int_{-\infty }^{+\infty }%
\frac{dk}{2\pi k}\frac{e^{+ikx}}{F\left( k\right) }\frac{\omega(k) \sinh \left(
ky\right) }{\sinh \left( \frac{kW}{2}\right) },  \label{P23}
\end{equation}
\begin{equation}
\mathcal{P}\left( \vec{r}\right) =\frac{I\tan \left( \theta\right) }{iD_{\parallel}}\int_{-\infty }^{+\infty }\frac{dk}{2\pi }\frac{%
e^{+ikx}}{F\left( k\right) }\frac{\cosh \left( \omega(k) y\right) }{\sinh
\left( \frac{\omega(k) W}{2}\right) },  \label{P24}
\end{equation}
where $\rho_c = 1/\sigma_{||}$, $F\left( k\right) =\tan ^{2}(\theta)\: k\coth \left( \frac{\omega(k)
W}{2}\right) +\omega(k) \coth \left( \frac{kW}{2}\right)$, and $\omega \left( k\right) =\sqrt{k^{2}+\ell_{\mathrm{v}}^{-2}}.$

Hence, the nonlocal resistance  can be obtained from $R_{\mathrm{NL}}\left( x\right) =\left[
\phi \left( x,-W/2\right) -\phi \left( x,W/2\right) \right] /I.$
Substituting Eq. (\ref{P23}) yields:~\cite{PhysRevB.94.121408}
\begin{equation}
R_{\mathrm{NL}}\left( x\right) =2\rho _{\mathrm{c}}\int_{-\infty }^{+\infty }%
\frac{dk}{2\pi k}\: e^{+ikx}\frac{\omega(k)}{F\left( k\right) }.  \label{P26}
\end{equation}
For $\theta=0$, the nonlocal resistance reduces to the ohmic contribution:
\begin{equation}
 R_{\mathrm{NL}} ^{0} \left( x\right) = \frac{2\rho_c}{\pi} \mathrm{ln} \left| \coth \left( \frac{\pi x }{2W} \right)  \right|.\label{eq:ohm}
\end{equation}
Fig.~\ref{NLR} shows the results of numerically integrating
Eq. (\ref{P26}). At a fixed distance $x=1 \:  \mu\mathrm{m}$ away from the current injection point, Fig.~\ref{NLR}(b) shows the nonlocal resistance $R_{\mathrm{NL}}$ against the chemical potential $\mu$, for different values of the pseudo-magnetic field, $\mathcal{B}_s$. The nonlocal resistance arising from the combined effect of VHE and inverse VHE is enhanced at low doping. Panel (c) in Fig.~\ref{NLR} shows the dependence of the nonlocal resistance $R_{\mathrm{NL}}(x)$ with $x/L_{\mathrm{v}}$ at fixed chemical potential $\mu =0.1$ eV. Nonuniform strain enhances the nonlocal resistance relative to its ohmic value. At large $|x|$, and for  $W \ll  \ell_{\mathrm{v}}$, $ R_{\mathrm{NL}}$ decays according  to:
\begin{equation}
R_{\mathrm{NL}}\left( x\right)  =  \rho_c  \frac{W}{2L_{\mathrm{v}}} \frac{\tan^2(\theta)}{1+\tan^2(\theta)}e^{- |x|/ L_{\mathrm{v}} },\label{eq:asym}
\end{equation}
which agrees well with the numerical results for $|x| \gg \ell_{\mathrm{v}}\gg W$ (cf. Fig.~\ref{NLR}, showing that the ohmic contribution, cf. Eq.~\eqref{eq:ohm} is also much smaller in this limit). Note that, in this regime, the decay is controlled by $L_\mathrm{v}$ rather than the length scale $\ell_\mathrm{v}$ introduced in Eq.~\eqref{VDL}. This can be understand from the fact that Eq.~\eqref{P26} is obtained by solving the coupled diffusion and Laplace equations, which takes into account the buildup of electrostatic potential (due to the inverse valley Hall effect) along the channel. The latter
modifies the decay of $R_{\mathrm{NL}}\left( x\right)$ by effectively
replacing $\ell_\mathrm{v} = L_{\mathrm{v}} (1 + \omega^2_c \tau_D)^{1/2}$ by $L_\mathrm{v} = \sqrt{\mathcal{D}\tau_{\mathrm{v}}}$.

\section{Summary and outlook}\label{sec:sum}

We have developed a theory of the strain-induced classical valley Hall effect (VHE).  Specifically, using the quantum Boltzmann equation, we
have provided a microscopic derivation of  the equations governing the diffusion of valley polarization. The latter have been solved for a Hall bar device geometry with subject to nonuniform strain leading to uniform pseudo-magnetic field. The observable nonlocal resistance of the device has been obtained.  We found that for low doping the figure of merit of the VHE, namely the valley Hall angle, $\theta(T)$
can be of order unity even at room temperature. The
nonlocal resistance of the device decays exponentially.

Finally, it is interesting to consider the effect of a strain configuration leading to a slowly varying (on the scale of the Fermi wavelength) pseudo-magnetic field. The equations derived here are also applicable in this case,
with the caveat that in such a case $\omega_c \tau_D$ becomes space dependent. This complicates the solution of the diffusion equation, Eq.~\eqref{P6}, as the source term on the right-hand side $S(\vec{r})$ will not be a boundary term.  In addition, the diffusion coefficient $D_{\parallel} = D_{\parallel}(\vec{r})$ is now a function of  the position in the device. However, qualitative, one can still expect a nonlocal signal to exist even if the sing of the pseudo-magnetic field fluctuates in space because the nonlocal resistance depends quadratically on the valley Hall angle $\theta \sim \omega_c \tau_D$  as it arises from the combination of the direct and inverse valley Hall effects.
\acknowledgements
This work is supported by the Ministry of Science and Technology (Taiwan)
under contract number NSC 102- 2112-M-007-024-MY5, and Taiwan's National Center of Theoretical Sciences (NCTS). We thank F. Guinea, A. Kaverzin, and J. Song for useful discussion.

\appendix*
\section{Temperature dependent conductivities}

In this Appendix, we provide the expressions  to compute
the charge ($\sigma_{\parallel}$) and spin Hall ($\sigma_{\perp}$)
conductivities at temperature $T > 0$:
\begin{align}
\sigma_{\parallel}(T)&=\frac{e^{2}}{2\pi} \int
d\epsilon \, |\epsilon| \frac{\tau_{\mathrm{D}}  \left[ -\partial_{\mu }n^{0}\left(\epsilon-\mu \right) \right]}{\left( 1+\omega_{c}^{2}\tau _{\mathrm{D}
}^{2}\right)},\\
\sigma_{\perp}(T)&=\frac{e^{2}}{2\pi}\int d\epsilon\:  |\epsilon|\,  \frac{\omega _{c}\tau^2_{\mathrm{D}} \left[ -\partial_{\mu }n^{0}\left( \epsilon-\mu
\right) \right] }{\left( 1+\omega
_{c}^{2}\tau _{\mathrm{D}}^{2}\right) }.
\end{align}
where both  $\tau_D$ and $\omega_c$ are energy (i.e. Fermi momentum) dependent and $n^0(\epsilon) = \left[e^{\epsilon/k_B T} +1 \right]^{-1}$ is the Fermi-Dirac distribution. Notice that for $T \to 0$, we recover Eq.~\eqref{eq:cond}.

\bibliography{reference}

\end{document}